# Integrable Quantum Mappings

H.W. Capel and F. W. Nijhoff[†]

ABSTRACT. We discuss the canonical structure of a class of integrable quantum mappings, i.e. iterative canonical transformations that can be interpreted as a discrete dynamical system. As particular examples we consider quantum mappings associated with the lattice analogues of the KdV and MKdV equations. These mappings possess a non-ultralocal quantum Yang-Baxter structure leading to the existence of commuting families of exact quantum invariants. We derive the associated quantum Miura transformations between these mappings and the corresponding quantum bi-Hamiltonian structure.

## 1. Introduction

In a number of recent papers the quantization of *discrete* integrable models has been investigated, [1]-[6]. In these systems not only the spatial dimension, but also the time-variable is discrete. The classical counterparts of these systems, integrable lattice equations which are described by partial difference equations, were constructed and studied already some time ago, cf. e.g. [7]-[9]. More recently, a lot of insight has been gained into special clases of solutions, such as soliton solutions, similarity solutions leading to lattice and discrete Painlevé equations, cf. [10] and also [11], and pole solutions leading to a discretized Calogero-Moser model (see the other contribution of the second author to this volume). In [12], cf. also [13], periodic initial value problems on the lattice have been studied, leading to finite-dimensional reductions of the lattice equations which are *integrable mappings*, i.e. dynamical systems with a discrete-time evolution which can be integrated in the sense of Liouville, cf. [14]. This means that the discrete time-flow is the iterate of a canonical transformation, preserving a suitable symplectic structure, leading to a complete set of invariants which are in involution with respect to this symplectic form. As in the continuous-time situation one can then linearize the discrete-time flow on a hypertorus which is the intersection of the level sets of the invariants, [15, 16]. Integrable mappings have been considered from slightly different perspectives also in the recent literature, cf. e.g. [17]-[19].

In this article we focuss on the quantization of the mapping reductions of the lattice systems. A theory of *integrable quantum mappings* of Korteweg-de Vries (KdV) type was proposed first in [1], and elaborated further in [3]-[5]. The quantization of integrable mappings of KdV type and their generalizations is based on the existence of a non-ultralocal quantum Yang-Baxter (YB) structure. The central ingredients are Yang-Baxter matrices of Yangian (or trigonometric type) and

1991 *Mathematics Subject Classification*. Primary 58F07, 39A10; Secondary 70F10.

† Supported by the Alexander von Humboldt Foundation.

a zero-curvature structure given by a discrete-time ZS (Zakharov-Shabat) system of the form

$$(1.1) \qquad L'_n(\lambda) M_n(\lambda) \ = \ M_{n+1}(\lambda) L_n(\lambda) \ ,$$

which holds both on the classical as well as on the quantum level. In eq. (1.1) $\lambda$ is a spectral parameter, $L_n$ is the lattice translation operator at site $n$, and the prime denotes the discrete time-shift corresponding to a translation in the second lattice direction. As $L$ and $M$, in the quantum case, depend on operators, the question of operator ordering becomes important. Throughout this paper we impose in the quantum case as a normal order the order which is induced by the lattice enumeration, with $n$ increasing from the right to the left. Finite-dimensional mappings are obtained from (1.1) imposing a periodicity condition

$$(1.2) \qquad L_n(\lambda) \ = \ L_{n+P}(\lambda) \ , \ M_n(\lambda) \ = \ M_{n+P}(\lambda)$$

for some $P \in N$.

As a consequence of the ZS system (1.1) we have on the classical level a complete family of invariants of the mapping , namely by introducing the associated monodromy matrix $T(\lambda)$, obtained by gluing the elementary translation matrices $L_j$ along a line connecting the sites 1 and $P+1$ over one period $P$, namely

$$(1.3) \qquad T(\lambda) \equiv \overset{\leftarrow}{\underset{n=1}{\overset{P}{\prod}}} L_n(\lambda) \ .$$

In the classical case the traces of powers of the monodromy matrix are invariant under the mapping as a consequence of

$$(1.4) \qquad T'(\lambda) = M_{P+1}(\lambda) T(\lambda) M_1^{-1}(\lambda)$$

and the periodicity condition $M_{P+1} = M_1$, thus leading to a sufficent number of invariants which are obtained by expanding the traces in powers of the spectral parameter $\lambda$. The involution property of the classical invariants follows from the Poisson bracket

$$(1.5) \qquad \{tr\, T(\lambda), tr\, T(\mu)\} = 0$$

which can be derived from the non-ultralocal classical YB structure, cf. e.g. [5].

For the quantum mappings we use the $R$-matrix structure of [1, 3, 4] which is the quantum analogue of the non-ultralocal Poisson structure. In the continuous-time case such a novel quantization scheme was proposed in ref. [20], in connection with the quantum Toda theory. Similar structures with continuous time flow have been introduced also for the quantum Wess-Zumino-Novikov-Witten (WZNW) theory with discrete spatial variable, cf. [21]. When considering discrete-time flows some interesting new features arise. The conventional point of view, that the $M$-part of the Lax equations does not need to be considered explicitly in order to construct quantum invariants, is no longer true. In fact, in the discrete case one needs to establish the complete quantum algebra, containing commutation relations between the $L$-operators as well as between the $L$- and the $M$-operators, and between the $M$-operators themselves. As a consequence we find, in the quantum mappings under consideration, that a commuting family of quantum invariants is given by

$$\tau(\lambda) = tr(T(\lambda)K(\lambda)) \ , \tag{1.6}$$

where $K(\lambda)$ is suitable numerical matrix, obeying the reflection algebra relations, cf. also [22]. Thus we have

$$\tau'(\lambda) = \tau(\lambda) \ , \ [\tau(\lambda), \tau(\mu)] = 0 \ . \tag{1.7}$$

The matrix $K(\lambda)$ can be explicitely calculated using the intertwining relations between the $M$-matrix and the monodromy matrix, and for the KdV and MKdV mappings the exact quantum invariants will be explicitely given.

In this paper we concentrate on mappings coming from the lattice KdV and lattice modified KdV (MKdV) systems. Classically, these mappings will be derived on the basis of a discrete variational principle, whereas their quantum analogues arise from the action of a unitary operator on the quantum phase space. The integrability of these mappings, together with the construction of exact quantum invariants, is assessed on the basis of the non-ultralocal YB structure of the type described above. As an additional new result, we will derive the quantum Miura transformation between the KdV and MKdV mappings. This transformation (which is the natural analogue of the continuous Miura transformation, leading e.g. to the classical bi-Hamiltonian structure) can be expressed as a gauge transformation relating the corresponding Lax matrices $L_n(\lambda)$. On the basis of this Miura transformation, a double commutation structure for the quantum discrete-time Volterra (Kac-van Moerbeke) system will be derived.

## 2. Discrete Action and Generating Operator

In this section, we introduce the mappings of KdV and MKdV type via an action principle. This action principle ensures automatically the symplectic property of the mappings, i.e. the presence of a Poisson bracket structure which is invariant under the mapping. In the quantum case the Poisson brackets between dynamical variables are replaced by commutators of quantum operators, and in that case we will specify the unitary operator that generates the quantum mapping as a canonical transformation.

Let us start from the following action

$$S = \sum_{n,m \in Z} [-u_{n,m+1}(u_{n,m} - u_{n+1,m+1}) + \Phi(u_{n,m} - u_{n+1,m+1})] \ , \tag{2.1}$$

where $\Phi$ is some 'potential' function. The action is a functional of the lattice field $u_{n,m}$, which is a function of the sites $(n,m)$ of a two-dimensional lattice. The variational principle that $S$ is invariant under infinitesimal variations of the lattice field $u_{n,m}$, i.e. $\delta S/\delta u_{n,m} = 0$, yields a partial difference equation in terms of the 'difference field' $v_{n,m} = u_{n,m} - u_{n+1,m+1}$, namely

$$v_{n+1,m} - v_{n,m+1} + (D\Phi)(v_{n,m}) - (D\Phi)(v_{n+1,m+1}) = 0 \ , \tag{2.2}$$

in which $D\Phi$ denotes the derivative of $\Phi$ with respect to its argument[1]. In eq. (2.2) the function $\Phi$ can, in principle, still be an arbitrary function, but we shall be mainly concerned with two cases, associated with the KdV and MKdV lattice, cf. [12], for which this equation becomes exactly integrable, respectively

$$\Phi(v) = \epsilon\delta\log(\epsilon + v) ,  \tag{2.3}$$

$$\Phi(v) = \int_{v+\sigma}^{v+\rho} \log(1+e^{\xi})d\xi = Li_2(-e^{v+\sigma}) - Li_2(-e^{v+\rho}) . \tag{2.4}$$

---

[1] As stated above, we reserve the prime ′ for the notation of a discrete time-shift.

In eqs. (2.3) and (2.4) $\epsilon, \delta, \rho$ and $\sigma$ are free parameters, and $Li_2$ is the (Euler) dilogarithm function. The partial difference equation (2.2) is a lattice version of the KdV equation for the choice (2.3), and a lattice version of the MKdV equation for the choice (2.4), cf. [13]. It is interesting to note that the dilogarithm seems also to have appeared recently in the quantization of the sine-Gordon model on the lattice, [23]. Appropriate continuum limits relating these difference equations to the corresponding continuum equations can be found e.g. in [24].

Going from the partial difference equation (2.2) to an (integrable) mapping, i.e. a finite-dimensional discrete-time system, is done by taking for instance the coordinate $m$ as our discrete time-variable and to consider the shift in $m$, $m \mapsto m+1$ as the time-step update in the mapping. Thus, choosing inital conditions along the sites of a 'staircase' consisting of alternating horizontal and vertical links, i.e. $u_{j,j} =: u_{2j}$, $u_{j+1,j} =: u_{2j+1}$, we can label the updates after one iteration by putting $u'_{2j} = u_{j,j+1}$, resp. $u'_{2j+1} = u_{j+1,j+1} = u_{2j+2}$, which we can then calculate by using the partial difference equation (2.2). Finite-dimenaional mappings are easily obtained by imposing periodic boundary conditions along the staircase $u_{n+P,m+P} = u_{n,m} \Rightarrow u_j = u_{j+2P}$. Clearly, the solution of eq. (2.2) will then satisfy the same boundary conditions. The mapping associated with (2.2) can be formulated in terms of the new reduced variables

$$x_j = u_{2j} - u_{2j+2} , \quad \sum_{j=1}^{P} x_j = 0 , \tag{2.5}$$

and is given by

$$x''_j = x_{j+1} + (D\Phi)(x'_j) - (D\Phi)(x'_{j+1}) , \tag{2.6}$$

In the presence of the periodic boundary conditions we have a reduced action which is the sum over all iterates of a discrete-time Lagrangian $L(x, x')$ which is given by

$$L(x, x') = \sum_{n=1}^{P-1} \frac{1}{2}\left(x'_n - \sum_{j=1}^{n} x_j\right)^2 - W(x) , \tag{2.7}$$

$$W(x) = \sum_{n=1}^{P-1}\left[\frac{1}{2}x_n^2 + \frac{1}{2}\left(\sum_{j=1}^{n} x_j\right)^2 - \Phi(x_n)\right] - \Phi\left(-\sum_{j=1}^{P-1} x_j\right) , \tag{2.8}$$

where $x$ is shorthand for $(x_1, ..., x_{P-1})$ and in which the $x_n$ are varied independently. Although the original action (2.1) is varied with respect to the variables $u_{n,m}$, it is easily verified that, varying here with respect to $x_n$, the Lagrange equations

$$\frac{\partial L}{\partial x'_n} + \left(\frac{\partial L}{\partial x_n}\right)' = 0 \tag{2.9}$$

yield the proper discrete-time equations of motion (2.2), so that the $x_n$ can be used as the canonical variables.

To obtain the generating function of the canonical transformation, it is convenient to choose the following special form of the Legendre transformation

$$(2.10) \qquad H(x, y') = \sum_{n=1}^{P-1} y'_n (x'_n - \sum_{j=1}^{n} x_j) - L(x, x') \quad , \quad y'_n \equiv \frac{\partial L}{\partial x'_n} \ .$$

By infinitesimal variation with respect to the variables $x_n$ and $y'_n$ of (2.10), and using (2.9), we obtain

$$(2.11) \qquad y_n - \sum_{j=n}^{P-1} y'_j = \frac{\partial H}{\partial x_n} \ ,$$

$$(2.12) \qquad x'_n - \sum_{j=1}^{n} x_j = \frac{\partial H}{\partial y'_n} \ ,$$

($n = 1, ..., P-1$), which can be interpreted as the discrete-time Hamilton equations.

From eqs. (2.11),(2.12) it is easily established that the variables $x_n$ and $y_n$ are canonical variables. It follows that the following Poisson brackets

$$(2.13) \qquad \{x_n, y_m\} = \delta_{n,m} \quad , \quad \{x_n, x_m\} = \{y_n, y_m\} = 0 \ ,$$

are preserved under the mapping given by (2.11),(2.12). The discrete-time 'hamiltonian' in (2.10) has the form

$$(2.14) \qquad H(x, y') = T(y') + W(x) \quad , \quad T(y) = \sum_{n=1}^{P-1} \frac{1}{2} y_n^2 \ ,$$

where $W(x)$ is given by eq. (2.8).

The quantization of discrete-time modes with hamiltonian (2.14) and Poisson brackets (2.13) is obtained by the straightforward quantization prescription $\{\cdot, \cdot\} \to \frac{1}{i\hbar}[\cdot, \cdot]$, replacing the canonical coordinates and we replace $x_n, y_n$ by hermitian quantum operators $x_n, y_n$ acting on a well-defined Hilbert space. Thus, we obtain for $x_n, y_n$ the Heisenberg algebra

$$(2.15) \qquad [x_n, y_m] = i\hbar \delta_{n,m} \quad , \quad [x_n, x_m] = [y_n, y_m] = 0 \ .$$

Now, as for the quantum version of the mapping, we first note that as a consequence of the splitting of $H$ into $T + W$ and the $x, y$ being canonically conjugate, the form of the mapping need not to be modified in the transition from the classical to the quantum case. Hence eqs. (2.11),(2.12) are still valid in terms of the quantum variables $x_n, y_n$. Secondly, the splitting of $H$ suggests directly the form of the unitary operator that generates the quantum mapping, Due to the presence of the sums on the left-hand side of eqs. (2.11), (2.12) we have

$$(2.16) \qquad U = e^{\frac{i}{\hbar} W} \left( \overleftarrow{\prod_{i=1}^{P-1}} \overleftarrow{\prod_{j=i+1}^{P-1}} e^{\frac{i}{\hbar} x_i y_j} \right) e^{\frac{i}{\hbar} T} \ ,$$

in which $W = W(x), T = T(y)$, and the exponential factors are ordered in lexicographic order form the right to the left. In fact, with $U$ given by eq. (2.16) the transformations

$$\text{(2.17)} \qquad x_n \mapsto x'_n = U x_n U^\dagger \ , \quad y_n \mapsto y'_n = U y_n U^\dagger$$

yield exactly the quantum mapping provided by the quantization of eqs. (2.11),(2.12). Eq. (2.16), together with (2.17), demonstrates the fact that the mappings introduced in this section are provided by a unitary transformation in the quantum phase space. A direct construction of the unitary operators defining the evolution for the lattice sine-Gordon system in terms of the quantum $L$-matrices was given in [6].

## 3. Non-ultralocal Yang-Baxter structure

We are interested in the canonical structure of discrete-time integrable systems, i.e. systems for which the time evolution is given by an iteration of mappings. If the mapping contains quantum operators, the commutation relations with the monodromy or Lax matrices become nontrivial and it is not a priori clear in this case that the YB structure is preserved. Furthermore, the traces of powers of the monodromy matrix are no longer trivially invariant as the cyclic property of the traces is no longer true for operator-valued arguments. To deal with these new features, it is necessary to take the $M$-part of the zero-curvature system (1.1) also into account, and investigate the *full* quantum structure involved in these systems, consisting of commutation relations between the $L$-part as well as of the $M$-part of the Lax pair. A proposal for such combined system was presented first in [3]. We will outline this novel YB structure here, and in the next section we will present explicit examples of quantum mappings that were derived in the previous section, namely the quantum mappings of KdV and MKdV type, that will be shown to exhibit this combined temporal-spatial YB structure.

The general *quantum* $L$-operator $L_n(\lambda)$, at each site $n$ of a one-dimensional lattice, is a matrix whose entries are quantum operators (acting on some properly chosen Hilbert space). We consider operators $L_n(\lambda)$ that have only non-trivial commutation relations between themselves on the same and nearest-neighbour sites, namely as follows

$$\text{(3.1)} \qquad \begin{aligned} R^+_{12} L_{n,1} L_{n,2} &= L_{n,2} L_{n,1} R^-_{12} \\ L_{n+1,1} S^+_{12} L_{n,2} &= L_{n,2} L_{n+1,1} \ , \\ L_{n,1} L_{m,2} &= L_{m,2} L_{n,1} \ , \quad \mid n - m \mid \geq 2 \ . \end{aligned}$$

We adopt here the usual convention that the subscripts $1, 2, \cdots$ denote the factors in a matricial tensor product. For example, in eq. (3.1), the subscripts $\alpha, \beta = 1, 2, \cdots$ for the operator matrices $L_{n,\alpha}$ denote the corresponding factor on which this $L_n$ acts (acting trivially on the other factors), i.e. $L_{n,1} = L_n(\lambda_1) \otimes 1$, $L_{n,2} = 1 \otimes L_n(\lambda_2)$. For notational convenience we suppress the explicit dependence on the spectral parameter $\lambda = \lambda_1$ respectively $\lambda = \lambda_2$, assuming always that each value accompanies its respective factor in the tensor product. The relations (3.1) were first proposed in [20] for the quantum Toda theory. In [21] a similar structure is referred to as the Kac-Moody algebra on the lattice.

The compatibility relations of the equations (3.1) lead to the following consistency conditions on $R^\pm$ and $S$

$$(3.2) \quad R^\pm_{12} R^\pm_{13} R^\pm_{23} = R^\pm_{23} R^\pm_{13} R^\pm_{12} ,$$

$$(3.3) \quad R^\pm_{23} S^\pm_{12} S^\pm_{13} = S^\pm_{13} S^\pm_{12} R^\pm_{23} ,$$

where $S^+_{12} = S^-_{21}$. Eq. (3.2) is the quantum YB equation (QYBE's) for $R^\pm$ coupled with an additional equation (3.3) for $S^\pm$. In order to establish that the structure given by the commutation relations (3.1) allows for suitable commutation relations for the monodromy matrix, we need to impose in addition to (3.2),(3.3) that

$$(3.4) \quad R^\pm_{12} S^\pm_{12} = S^\mp_{12} R^\mp_{12} .$$

Taking into account the periodic boundary conditions the following commutation relations for the monodromy matrix can be obtained

$$(3.5) \quad R^+_{12} T_1 S^+_{12} T_2 = T_2 S^-_{12} T_1 R^-_{12} ,$$

where $T_1 = T(\lambda_1) \otimes 1$, and $T_2 = 1 \otimes T(\lambda_2)$.

Let us now consider the discrete-time part of the zero-curvature system (1.1). The YB structure is now extended, by supplying in addition to eqs. (3.1) the following system of equations

$$(3.6) \quad \begin{aligned} M_{n+1,1} S^+_{12} L_{n,2} &= L_{n,2} M_{n+1,1} , \\ L'_{n,2} S^-_{12} M_{n,1} &= M_{n,1} L'_{n,2}, \end{aligned}$$

as well as

$$(3.7) \quad \begin{aligned} R^+_{12} M_{n,1} M_{n,2} &= M_{n,2} M_{n,1} R^-_{12} , \\ M'_{n,1} S^+_{12} M_{n,2} &= M_{n,2} M'_{n,1} . \end{aligned}$$

On top of (3.6) and (3.7) we need to give a number of trivial commutation relations, namely

$$(3.8) \quad [M_{n,1}, L_{m,2}] = [M_{n,1}, M_{m,2}] = [M_{n+1,1}, L'_{m,2}] = 0 \ , \ |n - m| \geq 2 .$$

We shall not specify other commutation relations, as they do not belong to the YB structure. That does not mean that there are no other nontrivial commutation relations. In fact, there are always number of model-dependent commutation relations, which, however, are not relevant to establish exact quantum invariants of the mapping.

The equations (3.6)-(3.8) ensure that the commutation relations (3.1) between the matrices $L_n$ are invariant under the quantum mapping given by (1.1). For the explicit examples, there are of course alternative ways of establishing the symplecticity of the mapping, like, for instance, the construction of the unitary operator as was done in section 2 for the KdV and MKdV mappings.

Having established the algebraic structure of the quantum mappings on a local (i.e. $n$-dependent) level, we then can 'integrate' this structure by considering the commutation relations for the monodromy matrix $T(\lambda)$. From (1.3) and the relations (3.1),(3.6) we are led to the following commutation relations between $\mathcal{M} \equiv M_{n=1}$ and the monodromy matrix $T$,

$$(3.9) \quad T_1 \mathcal{M}_1^{-1} S^+_{12} \mathcal{M}_2 = \mathcal{M}_2 S^-_{12} T_1 \mathcal{M}_1^{-1} .$$

Using (3.9) it is an easy excercise to show that the commutation relations (3.5) for the monodromy matrices are invariant under the mapping. Also using the relation (3.9) we can construct exact quantum invariants of the mapping in the form of the parameter-family (1.6). The matrix $K(\lambda)$ has to obey two conditions. The first condition is the reflection algebra relations

$$(3.10) \qquad K_1{}^{t_1}\left(({}^{t_1}S_{12}^-)^{-1}\right) K_2 R_{12}^+ = R_{12}^- K_2{}^{t_2}\left(({}^{t_2}S_{12}^+)^{-1}\right) K_1 ,$$

which one needs in order to establish the commutativity of the parameter-family $\tau(\lambda)$, as follows from the treatment of ref. [22], cf. also [5] for the details of the derivation. In eq. (3.10) we assume, of course, that $S_{12}^\pm$ and $R_{12}^\pm$ are invertible. The left superscripts ${}^{t_1}$ and ${}^{t_2}$ denote the matrix transpositions with respect to the corresponding factors 1 and 2 in the matricial tensor product. The second condition is the relation

$$(3.11) \qquad tr_1(P_{12} K_2 S_{12}^+) = 1_2 ,$$

where $P_{12}$ is the permutation operator satisfying $P_{12} A_1 = A_2 P_{12}$, $P_{12} = P_{21}$, $tr_1 P_{12} = 1_2$, (for matrices $A$ not depending on the spectral parameter) and where $tr_1$ denotes the trace over the first factor in the matricial tensor product. From (3.11) together with (3.9), one can establish that the family $\tau(\lambda)$ is actually *invariant* under the quantum mapping. The general solution of (3.11) is given by

$$(3.12) \qquad K_2 = tr_1\left\{P_{12}{}^{t_1}\left(({}^{t_1}S_{12}^+)^{-1}\right)\right\} ,$$

which can be shown to obey also the reflection relation (3.10) provided the $R$- and $S$-matrices obey the YB relations (3.2)- (3.4). Thus, we have an explicit construction of invariants for the quantum mappings, which we can apply to the special cases of the KdV and MKdV mappings.

## 4. The KdV and MKdV Mappings

Here we consider two examples of integrable quantum mappings coming from the lattice analogues of the KdV and MKdV equations.

a) The first example of a quantum mapping that exhibits the structure outlined above, is the mapping of the KdV type (i.e. eqs. (2.2) for the choice of potential given by (2.3)). From (2.11),(2.12) with $v_{2j} = x_j + \epsilon$, $v_{2j-1} = x_j - y_j - x_{j-1} + y_{j-1} + \epsilon$, cf. (2.10) together with (2.3), we obtain the following $(2P-2)$-dimensional generalization of the McMillan [25] mapping

$$(4.1)\ v'_{2j-1} = v_{2j} , \quad v'_{2j} = v_{2j+1} + \frac{\epsilon\delta}{v_{2j}} - \frac{\epsilon\delta}{v_{2j+2}} \qquad (j = 1,\cdots,P-1),$$

To obtain the YB structure it is worthwhile to note that eq. (4.1) arises as the compatibility conditions (1.1) with

$$(4.2) \quad L_j = V_{2j} V_{2j-1} , \quad M_j = \begin{pmatrix} a_j & 1 \\ \lambda_{2j} & 0 \end{pmatrix} , \quad V_i = \begin{pmatrix} v_i & 1 \\ \lambda_i & 0 \end{pmatrix} ,$$

in which $\lambda_{2j} = k^2 - q^2$, $\lambda_{2j+1} = k^2 - p^2$ and $\epsilon\delta = p^2 - q^2$. In fact, from (1.1) one obtains $a_j = v_{2j-1} - \epsilon\delta/v_{2j}$, as well as the mapping (4.1). In the quantum case the variables $v_j$ are hermitean operators with Heisenberg type of commutation relations, cf. (2.13),

$$(4.3) \qquad [v_j, v_{j'}] = i\hbar \left( \delta_{j,j'+1} - \delta_{j+1,j'} \right) .$$

The symplecticity of the mapping (4.1) with respect to the commutation relations (4.3) can be checked by direct computation.

The special solution of the quantum relations (3.2)-(3.4), which constitutes the $R, S$-matrix structure for the quantum mapping (4.1), together with the commutation relation (4.3), is given by

$$(4.4) \qquad R_{12}^- = 1 \otimes 1 + i\hbar \frac{P_{12}}{k_1^2 - k_2^2} \quad , \quad R_{12}^+ = R_{12}^- - S_{12}^+ + S_{12}^- ,$$

$$S_{12}^+ = 1 \otimes 1 - \frac{i\hbar}{k_2^2 - q^2} F \otimes E \quad , \quad S_{12}^- = S_{21}^+ ,$$

in which the permutation operator $P_{12}$ and the matrices $E$ and $F$ are given by

$$P_{12} = E \otimes F + F \otimes E + EF \otimes FE + FE \otimes EF ,$$

$$(4.5) \qquad E = \begin{pmatrix} 0 & 1 \\ 0 & 0 \end{pmatrix} \quad , \quad F = \begin{pmatrix} 0 & 0 \\ 1 & 0 \end{pmatrix} .$$

Calculating the matrix $K$ using eq. (3.12) for the KdV quantum mapping, we find

$$(4.6) \qquad K(\lambda) = 1 + \frac{i\hbar}{\lambda} FE \quad , \quad \lambda = k^2 - q^2 .$$

Thus, by expanding the operator $\tau$ of (1.6) using (4.6) in powers of the spectral parameter $k^2$, we can explicitely find the exact quantum invariants of the mapping (4.1).

b) The second example of a quantum mapping associated with the structure developed in the previous section, we consider

$$(4.7) \qquad \varphi'_{2n-1} = \varphi_{2n} \quad , \quad e^{\varphi'_{2n}} = e^{\gamma_{n+1}} e^{\varphi_{2n+1}} e^{-\gamma_n} ,$$

where

$$e^{\gamma_n} = \frac{(p_{2n} - r) + (p_{2n-1} + r)e^{\varphi_{2n}}}{(p_{2n-1} - r) + (p_{2n} + r)e^{\varphi_{2n}}} .$$

Eq. (4.7) is the mapping (2.2) for the potential (2.4), and which is the quantum mapping associated with the lattice MKdV equation, cf. [**12, 13**], and it arises as the compatibility condition of the zero-curvature system (1.1) with $L_n = V_{2n} V_{2n-1}$, and

$$(4.8) \qquad V_n = \Lambda_n \begin{pmatrix} 1 & 0 \\ 0 & e^{\varphi_n} \end{pmatrix} \quad , \quad \Lambda_n = \begin{pmatrix} p_n - r & 1 \\ \lambda & p_n + r \end{pmatrix} .$$

in which $\lambda = k^2 - r^2$, $p_{2n-1} = p$, $p_{2n} = q$, and

$$(4.9) \qquad M_n = \Lambda_{2n} \begin{pmatrix} 1 & 0 \\ 0 & e^{\gamma_n} \end{pmatrix} .$$

In fact, working out (1.1) with (4.8) and (4.9), one finds $\exp(\gamma_n)$ and the mapping (4.7). The quantum commutation relations are, (with the abbreviation $h = i\hbar$),

$$(4.10) \qquad e^{\varphi_n} e^{\varphi_{n+1}} = e^{-h} e^{\varphi_{n+1}} e^{\varphi_n} .$$

The YB structure for the MKdV mappings is given in terms of the following $R, S$-matrices. Introducing

(4.11)
$$R_{12}^-(\lambda_{12}) = \begin{pmatrix} e^{-h}\lambda_{12} - 1 & 0 & 0 & 0 \\ 0 & \lambda_{12} - 1 & e^{-h} - 1 & 0 \\ 0 & \lambda_{12}(e^{-h} - 1) & e^{-h}(\lambda_{12} - 1) & 0 \\ 0 & 0 & 0 & e^{-h}\lambda_{12} - 1 \end{pmatrix},$$

together with

(4.12)
$$S_{12} = 1 + (e^{-h} - 1)FE \otimes FE,$$

in which $\lambda_{12} = \lambda_1/\lambda_2$. Eq. (4.11) together with (4.12) yields a solution of the YB relations (3.2)-(3.4), namely

(4.13)
$$\begin{aligned} R_{12}^+ &= \Lambda_1\Lambda_2 R_{12}(\lambda_1/\lambda_2)(\Lambda_1\Lambda_2)^{-1}, \\ S_{12}^+ &= \Lambda_2 S_{12}\Lambda_2^{-1}, \quad S_{12}^- = S_{21}^+ = \Lambda_1 S_{21}\Lambda_1^{-1}. \end{aligned}$$

Calculating the matrix $K$ from eq. (3.12) in the present case we find

(4.14)
$$K(\lambda) = 1 + (e^h - 1)FE\Lambda_{2n}FE\Lambda_{2n}^{-1},$$

leading again to a commuting family of quantum invariants of the MKdV mapping (4.7) using the constructions of the previous section.

We note that the explicit expressions for the matrix $K(\lambda)$ in both the KdV and MKdV cases will lead to quantum corrections in the corresponding invariants which are visible only in the boundary terms coming from the contributions of the matrices $L_1$ and $L_{2P}$ at the beginning and the end of the staircase defining the monodromy matrix $T$.

## 5. Quantum Miura Transformation

So far, we have derived the same YB structure with the associated commutation relations (3.1)-(3.7) for the quantum KdV and MKdV mappings separately. However, we know that the lattice KdV and MKdV –much like their continuum counterparts– are related via a Miura transformation, cf. [12, 13]. Thus, it is natural to pursue the interrelation between these two models also for the quantum case in the mapping reduction. Most conveniently, this Miura transformation can be expressed as a gauge transformation

(5.1)
$$V_j^{MKDV} = G_j V_j^{KDV} G_{j-1}^{-1}, \quad G_j = \begin{pmatrix} \alpha_j & \beta_j \\ \gamma_j & 0 \end{pmatrix},$$

where $V_j^{MKDV}$ and $V_j^{KDV}$ are the matrices $V_j$ in the respective cases given in eq. (4.2) and (4.8) respectively. The entries of the gauge matrix $G_j$ is given by can be calculated as

(5.2)
$$\alpha_j = (p_j - r)\beta_{j-1}, \quad \gamma_j = (k^2 - r^2)\beta_{j-1}, \quad e^{\varphi_j} = \beta_j\beta_{j-2}^{-1}.$$

The Miura transformation coming from (5.1) reads

(5.3)
$$v_j = (p_j - r)\beta_{j-1}^{-1}\beta_{j-2} + (p_j + r)\beta_{j-1}^{-1}\beta_j.$$

["

interesting to note, that apparently for discrete systems in the quantum case, there exist different (Miura) transformations associated with different commutation systems for one and the same field. In this respect, a comparison with similar results for the continuous-time situation, i.e. e.g. the results of [30, 31], seems to indicate that in the discrete-time case some modifications arise due to the 'staggering' effect of the initial-value staircase. It would be of interest also to extend the above results to the Miura transformations that we derived in [32] for the mappings related to the so-called lattice Gel'fand-Dikii hierarchy and its associated quantization, cf [4]. We expect possible relations with discrete versions of $W$-algebras that were derived recently, cf. e.g. [33, 34].

# References

1. F.W. Nijhoff, H.W. Capel and V.G. Papageorgiou, *Integrable Quantum Mappings*, Phys. Rev. **A46** (1992), 2155–2158.
2. G.R.W. Quispel and F.W. Nijhoff, *Integrable Two-dimensional Quantum Mappings*, Phys. Lett. **161A** (1991), 419–422.
3. F.W. Nijhoff and H.W. Capel, *Integrable Quantum Mappings and Non-ultralocal Yang-Baxter Structures*, Phys. Lett. **A163** (1992), 49–56.
4. F.W. Nijhoff and H.W. Capel, *Integrability and Fusion Algebra for Quantum Mappings*, J. Phys. **A26** (1993), 6385–6407.
5. F.W. Nijhoff and H.W. Capel, *Quantization of Integrable Mappings*, Springer Lect. Notes Phys. **424**, (1993), 187–211.
6. L.D. Faddeev and A.Yu. Volkov, *Quantum Inverse Scattering Method on a Space-Time Lattice*, Theor. Math. Phys. **92** (1992), 837–842.
7. M.J. Ablowitz and F.J. Ladik, *On the Solution of a Class of Nonlinear Partial Difference Equations*, ibid. **57** (1977) 1–12; R. Hirota, *Nonlinear Partial Difference Equations I-III*, J. Phys. Soc. Japan **43** (1977) 1424–1433, 2074–2089; E. Date, M. Jimbo and T. Miwa, *Method for Generating Discrete Soliton Equations I-V*, J. Phys. Soc. Japan **51** (1982), 4116–4131, **52** (1983), 388–393, 761–771.
8. F.W. Nijhoff, G.R.W. Quispel and H.W. Capel, *Direct Linearization of Nonlinear Difference-Difference Equations*, Phys. lett. **97A** (1983), 125–128; G.R.W. Quispel, F.W. Nijhoff, H.W. Capel and J. van der Linden, *Linear Integral Equations and Nonlinear Difference-Difference Equations*, Physica **125A** (1984), 344–380.
9. F.W. Nijhoff, H.W. Capel, G.L. Wiersma and G.R.W. Quispel, *Bäcklund Transformations and Three-Dimensional Lattice Equations*, Phys. Lett. **105A** (1984), 267–272; F.W. Nijhoff, H.W. Capel and G.L. Wiersma, *Integrable Lattice Systems in Two and Three Dimensions*, Springer Lect. Notes Phys. **239** (1985), 263–302.
10. F.W. Nijhoff and V.G. Papageorgiou, *Similarity Reductions of Integrable Lattices and Discrete Analogues of the Painlevé II Equation*, Phys. Lett.**153A** (1991), 337–344.
11. B. Grammaticos and A. Ramani, *Discrete Painlevé Equations: Derivation and Properties*, in Applications of Analytic and Geometric Methods to Nonlinear Differential Equations, P.A. Clarkson (Ed.), NATO ASI Series C, vol. 413, (Kluwer Acad. Publ., Dordrecht, 1993), pp. 299–314.
12. V.G. Papageorgiou, F.W. Nijhoff and H.W. Capel, *Integrable Mappings and Nonlinear Integrable Lattice Equations*, Phys. Lett. **147A** (1990), 106–114.
13. H.W. Capel, F.W. Nijhoff and V.G. Papageorgiou, *Complete Integrability of Lagrangian Mappings and Lattices of KdV Type*, Phys. Lett. **155A** (1991), 377–387.
14. F.W. Nijhoff, V.G. Papageorgiou and H.W. Capel, *Integrable Time-Discrete Systems: Lattices and Mappings*, Springer Lecture Notes Math. **1510** (1992), 312–325.
15. A.P. Veselov, *Integrable Maps*, Russ. Math. Surv. **46** (1991), 1–51.
16. M. Bruschi, O. Ragnisco, P.M. Santini and G.-Z. Tu, *Integrable Symplectic Maps*, Physica **49D** (1991), 273–294.
17. G.R.W Quispel, J.A.G. Roberts and C.J. Thompson, *Integrable Mappings and Soliton Equations*, Phys. Lett. **A126** (1988), 419–421; ibid. II, Physica **D34** (1989), 183–192.
18. P.A. Deift, L.C. Li and C. Tomei, *Matrix Factorizations and Integrable Systems*, Commun.

Pure Appl. Math. **42** (1989), 443–521; J. Moser and A.P. Veselov, *Discrete Versions of Some Classical Integrable Systems and Factorization of Matrix Polynomials*, Commun. Math. Phys. **139** (1991), 217–243.
19. Yu. B. Suris, *Integrable Mappings of the Standard Type*, Funct. Anal. Appl. **23** (1989), 74–79; *Generalized Toda Lattices in Discrete Time*, Leningrad Math. J. **2** (1991), 339–352.
20. O. Babelon and L. Bonora, *Quantum Toda Theory*, Phys. Lett. **253B** (1991), 365–372.
21. A. Alekseev, L.D. Faddeev and M.A. Semenov-Tian-Shanskii, *Hidden Quantum Groups inside Kac-Moody Algebras*, Springer Lecture Notes Math. **1510** (1992), 148–158.
22. E.K. Sklyanin, *Boundary Conditions for Integrable Quantum Systems*, J. Phys. **A21** (1988), 2375–2389.
23. V.V. Bazhanov, private communication.
24. G. Wiersma and H.W. Capel, *Lattice Equations, Hierarchies and Hamiltonian Structures*, Physica **142A** (1987), 199–244; ibid. **149A** (1988), 49–74.
25. E.M. McMillan, in *Topics in Physics*, eds. W.E. Brittin and H. Odabasi, (Colorado Associated Univ. Press, Boulder, 1971), p. 219.
26. F.W. Nijhoff, *On a q-deformation of the Discrete Painlevé I Equation and q-Orthogonal Polynomials*, Lett. Math. Phys. **30** (1994) 327–336.
27. L.D. Faddeev and L.A. Tahktajan, *Liouville Model on the Lattice*, Springer Lect. Notes Phys. **246** (1986) 166–179.
28. O. Babelon, *Exchange Formula and Lattice Deformation of the Virasoro Algebra*, Phys. Lett. **238B** (1990), 234–241.
29. A.Yu. Volkov, *Miura transformation on the Lattice*, Theor. Math. Phys. **74** (1988) 96–99.
30. A.Yu. Volkov, *Quantum Volterra Model*, Phys. Lett. **A167** (1992), 345–355.
31. O. Babelon, *Liouville Theory on the Lattice and Universal Exchange Algebra for Bloch Waves*, Springer Lecture Notes Math. **1510** (1992) 159–175.
32. F.W. Nijhoff, H.W. Capel, G.R.W. Quispel and V.G. Papageorgiou, *The Lattice Gel'fand-Dikii Hierarchy*, Inverse Probl. **8** (1992) 597–621.
33. S.V. Kryukov and Ya.P. Pugay, *Lattice W-Algebras and Quantum Groups*, Preprint Landau-93-TMP-5, hep-th/9310154.
34. A. Belov and K.P. Chaltikian, *Lattice Analogues of W-Algebras and Classical Integrable Systems*, Phys. Lett. **309B** (1993) 268; *Lattice Analogue of the W-infinity Algebra and Discrete KP-Hierarchy*, **317B** (1993), 64–72.

INSTITUTE FOR THEORETICAL PHYSICS, UNIVERSITY OF AMSTERDAM, VALCKENIERSTRAAT 65, 1018 XE AMSTERDAM, THE NETHERLANDS

DEPARTMENT OF MATHEMATICS, UNIVERSITY OF PADERBORN, D-33095 PADERBORN, GERMANY

*E-mail address*: nijhoff@uni-paderborn.de